\documentclass[prd,twocolumn,showpacs,floatfix,amsmath,nofootinbib,amssymb,floatfix]{revtex4}
\usepackage{graphicx,color,dcolumn,booktabs,bm}
\usepackage{longtable,lscape}
\usepackage{txfonts}
\usepackage{overpic}
\usepackage{amssymb}
\usepackage{indentfirst}
\usepackage{feynmf}   
\usepackage{slashed}  
\usepackage{cases}
\usepackage{color}
\usepackage{multirow}
\usepackage{epstopdf}
\usepackage{graphicx,color,dcolumn,booktabs,bm}
\usepackage[colorlinks, citecolor=blue,anchorcolor=red,menucolor=red, linkcolor=red,filecolor=red,runcolor=red,urlcolor=blue,frenchlinks=red]{hyperref}

\graphicspath{{Figures/}} %
\begin{document}
\title{$X(5568)$ and its partner states}
\author{Yan-Rui Liu$^1$}
\email{yrliu@sdu.edu.cn} \affiliation{ $^1$School of Physics and Key
Laboratory of Particle Physics and Particle Irradiation (MOE),
Shandong University, Jinan 250100, China}
\author{Xiang Liu$^{2,3}$}
\email{xiangliu@lzu.edu.cn} \affiliation{
$^2$School of Physical Science and Technology, Lanzhou University, Lanzhou 730000, China\\
$^3$Research Center for Hadron and CSR Physics, Lanzhou University
and Institute of Modern Physics of CAS, Lanzhou 730000, China }
\author{Shi-Lin Zhu$^{4,5,6}$}
\email{zhusl@pku.edu.cn} \affiliation{ $^4$School of Physics and
State Key Laboratory of Nuclear Physics and Technology, Peking
University, Beijing 100871, China
\\
$^5$Collaborative Innovation Center of Quantum Matter, Beijing
100871, China
\\
$^6$Center of High Energy Physics, Peking University, Beijing
100871, China }

\date{\today}
\begin{abstract}
Stimulated by the recent observation of the $X(5568)$, we study the
$X(5568)$ and its partners under the tetraquark scenario. In the
framework of the color-magnetic interaction, we estimate the masses
of the partner states of the $X(5568)$ and discuss their decay
pattern, which provide valuable information on the future
experimental search of these states.
\end{abstract}

 \maketitle
\section{Introduction}\label{sec1}

Very recently, the D\O\, Collaboration observed a narrow structure
by analyzing the $B_s^0\pi^\pm$ invariant mass spectrum, which was
named as the $X(5568)$ \cite{D0:2016mwd}. The $X(5568)$ has a mass
$m=5567.8\pm2.9( \rm stat)^{+0.9}_{-1.9}(\rm syst)$ MeV and width
$\Gamma=21.9\pm6.4(\rm stat)^{+5.0}_{-2.5}(\rm syst)$ MeV
\cite{D0:2016mwd}. Its discovery mode indicates that the valence
quark component of the $X(5568)$ should be $\ell s\bar{\ell}\bar{b}$
($\ell=u$ or $d$). Thus, the $X(5568)$ is a good candidate of the
exotic tetraquark state. In the past decades, experimental search
for the exotic states and the corresponding theoretical
investigations have been an important research topic of hadron
physic (see recent review on the study of the exotic states in Ref.
\cite{Chen:2016qju}).

Under the general tetraquark scheme, there exist two possible
configurations for the $X(5568)$, i.e., the $\bar{B}K$ molecular
state and a compact tetraquark composed of a diquark and
anti-diquark. In Ref. \cite{Liu:2009uz}, Liu {\it et al.} studied
the $\bar{B}K$ interaction in the framework of chiral perturbative
theory. The authors found that the Weinberg-Tomozawa term for the
$\bar{B}K$ interaction in the isovector sector is zero
\cite{Liu:2009uz}, which shows that the attraction between $\bar{B}$
and $K$ is not strong. If assigning the $X(5568)$ as the $\bar{B}K$
molecular state, the binding energy of the $\bar{B}K$ system should
be around 200 MeV. In other words, this $\bar{B}K$ molecular state
is a deeply bound state. It is obvious that the study from Ref.
\cite{Liu:2009uz} does not support such a deeply bound $\bar{B}K$
molecular state scenario. We also notice a dynamical study of the
interaction between $\bar B$ and $K$ by exchanging the $\rho$ and
$\omega$ mesons \cite{Zhang:2006ix}. They did not find a bound state
solution for the S-wave $\bar BK$ system \cite{Zhang:2006ix}.

The observation of $X(5568)$ has inspired extensive discussions of
the possibility of a compact tetraquark state. Chen {\it et al.}
\cite{Chen:2016mqt} constructed $0^+$ and $1^+$ tetraquark current,
and adopted the QCD sum rule approach to calculate the corresponding
masses. Their result supports the $X(5568)$ as tetraquark state with
spin-parity quantum number $J^P=0^+$ or $1^+$. Additionally, the
charmed partner of the $X(5568)$ was predicted \cite{Chen:2016mqt}.
In Ref. \cite{Chen:2016mqt}, the authors considered the $X(5568)$ as
a tetraquark state with $J^P=0^+$ and calculated its mass
\cite{Agaev:2016mjb} and decay width \cite{Agaev:2016ijz}. Zanetti,
Nielsen and Khemchandani also adopted the QCD sum rule formalism to
study the $X(5568)$ as a tetraquark state with $J^P=0^+$
\cite{Zanetti:2016wjn}. A similar QCD sum rule study was performed
in Ref. \cite{Wang:2016mee}. Wang and Zhu applied the effective
Hamiltonian approach to calculate the mass spectrum of the
tetraquark state \cite{Wang:2016tsi}. They found that a S-wave
tetraquark with the quark component $[su][\bar b \bar d]$ and
$J^P=0^+$ lies 150 MeV higher than the $X(5568)$. There also exists
the discussion of the $X(5568)\to B_s\pi^+$ and $X(5616)\to
B_s^*\pi^+$ decay \cite{Xiao:2016mho}, where $X(5568)$ and $X(5616)$
was assumed as the S-wave ${B}\bar K$ and $B\bar K^*$ molecular
states, respectively. In Ref. \cite{Liu:2016xly}, Liu and Li explained the $X(5568)$ to be
the near threshold rescattering effect.

Twelve years ago, in order to explain its exotic decay modes, Liu
{\it et al.} once proposed a tetraquark structure for the $D_{sJ}(2632)$
signal observed by the SELEX collaboration \cite{Liu:2004kd}. The
$X(5568)$ corresponds to the bottom partner of the $D_{s,\bar{6}}$
or $D_{s,15}$ state there. The discovery of the $X(5568)$ signal
aroused our interest in the tetraquark candidates with four
different flavors again. If the $X(5568)$ is a tetraquark state, its
partner states within the same multiplet must also exist. In this
work, we mainly focus on the partner states of $X(5568)$ under the
tetraquark scenario. We will estimate the mass difference of these
partner states based on the color-magnetic interaction. Thus, the
present study hopefully provides valuable information on these
partners of the $X(5568)$. Experimental search for them can further
test the tetraquark assignment to the $X(5568)$.

At present, the LHCb collaboration \cite{LHCb} was unable to confirm this X(5568) signal
based on their own data. One should be aware of the different production
mechanism at Tevatron and LHCb, The huge amount of anti-quarks within the antiproton at Tevatron should
be helpful to the formation of this X(5568) with four-flavoured quarks
and anti-quarks if this signal really exists. Under the extreme case that this
X(5568) signal is real, it's highly probable that the current LHCb is unable
to observe it. If one tetraquark signal is observed experimentally, all the
other members within the same SU(3) flavor multiplet should also exist. Their
mass splittings can be estimated using the chromomagnetic interaction Hamiltonian.
In other words, the masses and decay modes of the partner states of the X(5568) can
be used to cross-check whether the X(5568) is a real resonance or not.
If these partner states are not observed, one should put a big question mark on the
existence of the X(5568) signal.

This paper is organized as follows. After the introduction, we
present the detailed formalism in Sec. \ref{sec2}. In Sec.
\ref{sec3}, we present the numerical results of the teatrquark
spectrum. In Sec. \ref{sec4}, we discuss their strong decay patterns
. The paper ends with a short discussion in Sec. \ref{sec5}.

\section{Framework}\label{sec2}

The low mass of $X(5568)$ suggests the existence of the tightly
bound four quark states in the presence of the heavy quark, which
may help stablize the tetraquark system. For such compact systems
confined within one MIT bag, the short range interaction should be
important. Here we adopt a simple chromomagnetic interaction model
containing the contact interaction only. The model Hamiltonian reads
\cite{Hyodo:2012pm}
\begin{eqnarray}
H&=&H_0+H_{CMI}\nonumber\\
&=&\sum_i
m_i+\sum_{i<j}\frac{C_{ij}}{m_im_j}\left(-\frac{3}{32}\right)\lambda_i\cdot\lambda_j\sigma_i\cdot\sigma_j,
\end{eqnarray}
where $\lambda$ ($\sigma$) is the generator for the color (spin)
symmetry and $C's$ are the parameters to be determined with known
hadrons.

In order to calculate the color-spin matrix elements, we explicitly
construct the $flavor\otimes color\otimes spin$ wave functions for
the tetraquark system $qq\bar{q}\bar{b}$. In discussing of the
$D_{sJ}(2632)$, we have obtained the flavor wave functions in
\cite{Liu:2004kd}. Here, we only need to extend formalism in
\cite{Liu:2004kd} and include the spin degree of freedom. We mainly
focus on the bottom mesons.

In the systems with one heavy quark, there exist degenerate spin
doublets in the heavy quark limit. The properties of the tetraquark
state is dominantly determined by the light quark cluster
$qq\bar{q}$. The flavor-color-spin wave functions of two quarks is
constrained by the Pauli principle. Therefore we mainly consider the
light diquark and do not assume any structure for the two antiquark
$\bar{q}\bar{Q}$. In other words, the tetraquark system is treated
as a triquark plus a heavy antiquark.

According to the diquark classification, we consider four cases for
the tetraquark structure $qq\bar{q}\bar{Q}$ in the present study:

(1) The representations for the $qq$ diquark are $\bar{3}_f$,
$\bar{3}_c$ and thus $S_{qq}=0,S_{qq\bar{q}}=1/2$. The final
tetraquark states form two flavor multiplets $3_f$ and $\bar{6}_f$.
In spin space, a degenerate $(0^+,1^+)$  doublet exists in the heavy
quark limit. In this case the diquark is a ``good'' diquark.

(2) The representations for the $qq$ diquark are $\bar{3}_f$, $6_c$
and thus $S_{qq}=1$, $S_{qq\bar{q}}=3/2$ or $1/2$. The final flavor
multiplets are $3_f$ and $\bar{6}_f$. Now in spin space, there are
two degenerate doublets $(1^+,2^+)$ and $(0^+,1^+)$. The color-spin
interaction for the two light quarks is weaker than the case (1) but
is still attractive. One can see this from the calculated values
below.

(3) The representation for the $qq$ diquark are $6_f$, $\bar{3}_c$
and thus $S_{qq}=1$, $S_{qq\bar{q}}=3/2$ or $1/2$. The final flavor
multiplets are $3_f$ and $15_f$. The spin doublets are the same as
the case (2). The color-spin interaction for the two light quarks is
weakly repulsive.

(4) The representation for the $qq$ diquark are $6_f$, $6_c$ and
thus $S_{qq}=0$, $S_{qq\bar{q}}=1/2$. The final flavor multiplets
are $3_f$ and $15_f$. There is one spin doublet $(0^+,1^+)$. The
color-spin interaction for the two light quarks is repulsive.

For the explicit flavor wave functions, one may consult Ref.
\cite{Liu:2004kd}. The color wave functions for $qq\bar{q}$ are
easily obtained with the replacement $u\to r$, $d\to g$, and $s\to
b$. The final color wave function for the tetraquark state depends
on the color state of $qq$. In the cases (1) and (3), it reads
\begin{eqnarray}
&&\frac{1}{2\sqrt3}[(rb\bar{b}-br\bar{b}-gr\bar{g}+rg\bar{g})\bar{r}+(gb\bar{b}-bg\bar{b}+gr\bar{r}-rg\bar{r})\bar{g}\nonumber\\
&&\qquad -(gb\bar{g}-bg\bar{g}+rb\bar{r}-br\bar{r})\bar{b}].
\end{eqnarray}
In the cases (2) and (4), one gets
\begin{eqnarray}
&&\frac{1}{2\sqrt6}[(rb\bar{b}+br\bar{b}+gr\bar{g}+rg\bar{g}+2rr\bar{r})\bar{r}
+(gb\bar{b}+bg\bar{b}+gr\bar{r}\nonumber\\
&&\quad +rg\bar{r}+2gg\bar{g})\bar{g}
+(gb\bar{g}+bg\bar{g}+rb\bar{r}+br\bar{r}+2bb\bar{b})\bar{b}].
\end{eqnarray}
These wave functions are constructed with the $SU(3)$ C.G.
coefficients given in Refs. \cite{deSwart:1963pdg,Kaeding:1995vq}.
The spin wave functions are easy to get and we do not show them
explicitly.

The $SU(3)$ flavor symmetry is actually violated by the unequal
quark mass. Such a breaking effect might be large in the multiquark
systems. As a result, there may exist the mixing between the states
with the same quantum numbers. In this work, we consider the
``ideal'' mixing. In the cases (1) and (2), the isospin one-half
states in $\bar{6}_f$ and $3_f$ mix and one gets the following
states
\begin{eqnarray}
&B^{\ell,+}=\frac{1}{\sqrt2}(ud-du)\bar{d}\bar{b},\quad B^{\ell,0}=\frac{1}{\sqrt2}(ud-du)\bar{u}\bar{b},&\nonumber\\
&B^{h,+}=\frac{1}{\sqrt2}(us-su)\bar{s}\bar{b},\quad
B^{h,-}=\frac{1}{\sqrt2}(ds-sd)\bar{s}\bar{b}.&
\end{eqnarray}
In the cases (3) and (4), the mixed isospin-half states are
\begin{eqnarray}
&B^{\prime\ell,+}=\frac{1}{\sqrt6}(ud\bar{d}+du\bar{d}+2uu\bar{u})\bar{b},&\nonumber\\ &B^{\prime\ell,0}=\frac{1}{\sqrt6}(ud\bar{u}+du\bar{u}+2dd\bar{d})\bar{b},&\nonumber\\
&B^{\prime h,+}=\frac{1}{\sqrt2}(us+su)\bar{s}\bar{b},\quad
B^{\prime h,0}=\frac{1}{\sqrt2}(ds+sd)\bar{s}\bar{b},
\end{eqnarray}
and the mixed isoscalar states are
\begin{eqnarray}
B^{\prime\ell}_s&=&\frac{1}{2}(us\bar{u}+su\bar{u}+ds\bar{d}+sd\bar{d})\bar{b},\nonumber\\
B^{\prime h}_s&=&ss\bar{s}\bar{b}.
\end{eqnarray}

\begin{table}[h!]
\begin{tabular}{ccc}\hline
Tetraquark& Quark Content&(Y,I)\\
$B_{\bar{s},\bar{6}}$&$ud\bar{s}\bar{b}$& $(\frac43,0)$\\
$B^\ell$&$\ell\ell\bar{\ell}\bar{b}$&$(\frac13,\frac12)$\\
$B^h$&$\ell s\bar{s}\bar{b}$&$(\frac13,\frac12)$\\
$B_{s,\bar{6}}$&$s\ell\bar{\ell}\bar{b}$&$(-\frac23,1)$\\
$B_{s,3}$&$s\ell\bar{\ell}\bar{b}$&$(-\frac23,0)$\\\hline
$B_{\bar{s},15}$&$\ell\ell\bar{s}\bar{b}$&$(\frac43,1)$\\
$B_{15}$&$\ell\ell\bar{\ell}\bar{b}$&$(\frac13,\frac32)$\\
$B^{\prime\ell}$&$\ell\ell\bar{\ell}\bar{b}$&$(\frac13,\frac12)$\\
$B^{\prime h}$&$\ell s\bar{s}\bar{b}$&$(\frac13,\frac12)$\\
$B_{s,15}$&$s\ell\bar{\ell}\bar{b}$&$(-\frac23,1)$\\
$B_s^{\prime\ell}$&$s\ell\bar{\ell}\bar{b}$&$(-\frac23,0)$\\
$B_s^{\prime h}$&$ss\bar{s}\bar{b}$&$(-\frac23,0)$\\
$B_{ss,15}$&$ss\bar{\ell}\bar{b}$&$(-\frac53,\frac12)$\\\hline
\end{tabular}
\caption{Tetraquarks, their quark contents, hypercharge and isospin,
where $\ell$ indicates $u$ or $d$ quark. The first five notations
are used in the cases (1) and (2). The other notations are used in
the cases (3) and (4). }\label{notations}
\end{table}

We adopt the notations in Ref. \cite{Liu:2004kd} for the other
tetraquarks and summarize them in Table \ref{notations}. With the
explicit expressions for the generators $\lambda$ and $\sigma$ and
the constructed wave functions, one can calculate the matrix element
$CMI=\langle H_{CMI}\rangle$.

\begin{table}[h!]
\begin{tabular}{cccccc|c||c}\hline
State&$(S_{qq\bar{q}},J)$&$\frac{C_{qq}}{m_q^2}$&$\frac{C_{qs}}{m_qm_s}$&$\frac{C_{\bar{q}\bar{b}}}{m_qm_b}$&$\frac{C_{\bar{s}\bar{b}}}{m_sm_b}$ & Mass&Charm case\\
\hline
$B_{\bar{s},\bar{6}}$&$(\frac12,1)$&8&&&$-\frac83$&5560(-)&2365(-)\\
(5700)&$(\frac12,0)$&8&&&8&5534(-)&2316(-)\\
\hline
$B^\ell$&$(\frac12,1)$&8&&$-\frac83$&&5417(-)&2224(-)\\
(5560)&$(\frac12,0)$&8&&8&&5403(-)&2181(-)\\\hline
$B^h$&$(\frac12,1)$&&8&&$-\frac83$&5776(-)&2582(-)\\
(5840)&$(\frac12,0)$&&8&&8&5750(-)&2533(-)\\\hline
$B_{s,\bar{6}}$, $B_{s,3}$&$(\frac12,1)$&&8&$-\frac83$&&5633(-)&2440(-)\\
(5700)&$(\frac12,0)$&&8&8&&5619(-)&2397(-)\\\hline
\end{tabular}
\caption{The obtained CMI's and estimated masses for the case (1) in
units of MeV. The symbol ``+'' (``-'') after the tetraquark mass
means that $H_{CMI}>0$ ($H_{CMI}<0$).}\label{CMI1}
\end{table}

\begin{table*}[h!]
\begin{tabular}{ccccccccccc|c||c}\hline
State&$(S_{qq\bar{q}},J)$&$\frac{C_{qq}}{m_q^2}$&$\frac{C_{qs}}{m_qm_s}$&$\frac{C_{q\bar{q}}}{m_q^2}$&$\frac{C_{q\bar{s}}}{m_qm_s}$&$\frac{C_{s\bar{s}}}{m_s^2}$&$\frac{C_{q\bar{b}}}{m_qm_b}$&$\frac{C_{s\bar{b}}}{m_sm_b}$&$\frac{C_{\bar{q}\bar{b}}}{m_qm_b}$&$\frac{C_{\bar{s}\bar{b}}}{m_sm_b}$
&Mass&Charm case\\
\hline $B_{\bar{s},\bar{6}}$
&$(\frac32,2)$&$\frac43$&&&$-\frac{20}{3}$&&$-\frac{20}{3}$&&&$\frac43$&5745(+)&2573(+)\\
(5700)&$(\frac32,1)$&$\frac43$&&&$-\frac{20}{3}$&&$\frac{100}{9}$&&&$-\frac{20}{9}$&5716(+)&2469(-)\\
&$(\frac12,1)$&$\frac43$&&&$\frac{40}{3}$&&$-\frac{40}{9}$&&&$-\frac49$&5569(-)&2390(-)\\
&$(\frac12,0)$&$\frac43$&&&$\frac{40}{3}$&&$\frac{40}{3}$&&&$\frac43$&5526(-)&2262(-)\\\hline
$B^\ell$
&$(\frac32,2)$&$\frac43$&&$-\frac{20}{3}$&&&$-\frac{20}{3}$&&$\frac{4}{3}$&&5671(+)&2498(+)\\
(5560)&$(\frac32,1)$&$\frac43$&&$-\frac{20}{3}$&&&$\frac{100}{9}$&&$-\frac{20}{9}$&&5637(+)&2392(+)\\
&$(\frac12,1)$&$\frac43$&&$\frac{40}{3}$&&&$-\frac{40}{9}$&&$-\frac{4}{9}$&&5301(-)&2122(-)\\
&$(\frac12,0)$&$\frac43$&&$\frac{40}{3}$&&&$\frac{40}{3}$&&$\frac{4}{3}$&&5260(-)&1995(-)\\\hline
$B^h$
&$(\frac32,2)$&&$\frac43$&&$-\frac{10}{3}$&$-\frac{10}{3}$&$-\frac{10}{3}$&$-\frac{10}{3}$&&$\frac43$&5945(+)&2772(+)\\
(5840)&$(\frac32,1)$&&$\frac43$&&$-\frac{10}{3}$&$-\frac{10}{3}$&$\frac{50}{9}$&$\frac{50}{9}$&&$-\frac{20}{9}$&5914(+)&2668(+)\\
&$(\frac12,1)$&&$\frac43$&&$\frac{20}{3}$&$\frac{20}{3}$&$-\frac{20}{9}$&$-\frac{20}{9}$&&$-\frac49$&5629(-)&2450(-)\\
&$(\frac12,0)$&&$\frac43$&&$\frac{20}{3}$&$\frac{20}{3}$&$\frac{20}{3}$&$\frac{20}{3}$&&$\frac43$&5585(-)&2322(-)\\\hline
$B_{s,\bar{6}}$, $B_{s,3}$
&$(\frac32,2)$&&$\frac43$&$-\frac{10}{3}$&$-\frac{10}{3}$&&$-\frac{10}{3}$&$-\frac{10}{3}$&$\frac{4}{3}$&&5793(+)&2619(+)\\
(5700)&$(\frac32,1)$&&$\frac43$&$-\frac{10}{3}$&$-\frac{10}{3}$&&$\frac{50}{9}$&$\frac{50}{9}$&$-\frac{20}{9}$&&5757(+)&2513(+)\\
&$(\frac12,1)$&&$\frac43$&$\frac{20}{3}$&$\frac{20}{3}$&&$-\frac{20}{9}$&$-\frac{20}{9}$&$-\frac{4}{9}$&&5518(-)&2339(-)\\
&$(\frac12,0)$&&$\frac43$&$\frac{20}{3}$&$\frac{20}{3}$&&$\frac{20}{3}$&$\frac{20}{3}$&$\frac{4}{3}$&&5475(-)&2212(-)\\
\hline
\end{tabular}
\caption{The obtained CMI's and estimated masses for the case (2) in
units of MeV. The symbol ``+'' (``-'') after the tetraquark mass
means that $H_{CMI}>0$ ($H_{CMI}<0$).}\label{CMI2}
\end{table*}

\begin{table*}[h!]
\begin{tabular}{cccccccccccc|c||c}\hline
State&$(S_{qq\bar{q}},J)$&$\frac{C_{qq}}{m_q^2}$&$\frac{C_{qs}}{m_qm_s}$&$\frac{C_{ss}}{m_s^2}$&$\frac{C_{q\bar{q}}}{m_q^2}$&$\frac{C_{q\bar{s}}}{m_qm_s}$&$\frac{C_{s\bar{s}}}{m_s^2}$&$\frac{C_{q\bar{b}}}{m_qm_b}$&$\frac{C_{s\bar{b}}}{m_sm_b}$&$\frac{C_{\bar{q}\bar{b}}}{m_qm_b}$&$\frac{C_{\bar{s}\bar{b}}}{m_sm_b}$
&Mass&Charm case\\
\hline $B_{\bar{s},15}$
&$(\frac32,2)$&$-\frac83$&&&&$-\frac83$&&$-\frac83$&&&$-\frac83$&5785(+)&2603(+)\\
(5700)&$(\frac32,1)$&$-\frac83$&&&&$-\frac83$&&$\frac{40}{9}$&&&$\frac{40}{9}$&5752(+)&2522(+)\\
&$(\frac12,1)$&$-\frac83$&&&&$\frac{16}{3}$&&$-\frac{16}{9}$&&&$\frac89$&5704(+)&2510(+)\\
&$(\frac12,0)$&$-\frac83$&&&&$\frac{16}{3}$&&$\frac{16}{3}$&&&$-\frac83$&5697(-)&2478(-)\\
\hline $B_{15}$, $B^{\prime\ell}$
&$(\frac32,2)$&$-\frac83$&&&$-\frac83$&&&$-\frac83$&&$-\frac83$&&5667(+)&2487(+)\\
(5560)&$(\frac32,1)$&$-\frac83$&&&$-\frac83$&&&$\frac{40}{9}$&&$\frac{40}{9}$&&5643(+)&2410(+)\\
&$(\frac12,1)$&$-\frac83$&&&$\frac{16}{3}$&&&$-\frac{16}{9}$&&$\frac{8}{9}$&&5514(-)&2319(-)\\
&$(\frac12,0)$&$-\frac83$&&&$\frac{16}{3}$&&&$\frac{16}{3}$&&$-\frac{8}{3}$&&5503(-)&2286(-)\\
\hline $B^{\prime h}$
&$(\frac32,2)$&&$-\frac83$&&&$-\frac43$&$-\frac43$&$-\frac43$&$-\frac43$&&$-\frac83$&5918(+)&2736(+)\\
(5840)&$(\frac32,1)$&&$-\frac83$&&&$-\frac43$&$-\frac43$&$\frac{20}{9}$&$\frac{20}{9}$&&$\frac{40}{9}$&5885(+)&2655(+)\\
&$(\frac12,1)$&&$-\frac83$&&&$\frac83$&$\frac83$&$-\frac89$&$-\frac89$&&$\frac{8}{9}$&5781(-)&2587(-)\\
&$(\frac12,0)$&&$-\frac83$&&&$\frac83$&$\frac83$&$\frac83$&$\frac83$&&$-\frac{8}{3}$&5774(-)&2556(-)\\
\hline $B_{s,15}$, $B^{\prime\ell}_s$
&$(\frac32,2)$&&$-\frac83$&&$-\frac43$&$-\frac43$&&$-\frac43$&$-\frac43$&$-\frac83$&&5769(+)&2589(+)\\
(5700)&$(\frac32,1)$&&$-\frac83$&&$-\frac43$&$-\frac43$&&$\frac{20}{9}$&$\frac{20}{9}$&$\frac{40}{9}$&&5744(+)&2512(+)\\
&$(\frac12,1)$&&$-\frac83$&&$\frac83$&$\frac83$&&$-\frac89$&$-\frac89$&$\frac{8}{9}$&&5654(-)&2459(-)\\
&$(\frac12,0)$&&$-\frac83$&&$\frac83$&$\frac83$&&$\frac83$&$\frac83$&$-\frac{8}{3}$&&5643(-)&2426(-)\\
\hline $B^{\prime h}_s$
&$(\frac32,2)$&&&$-\frac83$&&&$-\frac83$&&$-\frac83$&&$-\frac83$&6114(+)&2931(+)\\
(5980)&$(\frac32,1)$&&&$-\frac83$&&&$-\frac83$&&$\frac{40}{9}$&&$\frac{40}{9}$&6080(+)&2851(+)\\
&$(\frac12,1)$&&&$-\frac83$&&&$\frac{16}{3}$&&$-\frac{16}{9}$&&$\frac{8}{9}$&5922(-)&2727(-)\\
&$(\frac12,0)$&&&$-\frac83$&&&$\frac{16}{3}$&&$\frac{16}{3}$&&$-\frac{8}{3}$&5914(-)&2696(-)\\
\hline $B_{ss,15}$
&$(\frac32,2)$&&&$-\frac83$&&$-\frac83$&&&$-\frac83$&$-\frac83$&&5934(+)&2753(+)\\
(5840)&$(\frac32,1)$&&&$-\frac83$&&$-\frac83$&&&$\frac{40}{9}$&$\frac{40}{9}$&&5908(+)&2676(+)\\
&$(\frac12,1)$&&&$-\frac83$&&$\frac{16}{3}$&&&$-\frac{16}{9}$&$\frac{8}{9}$&&5857(+)&2662(+)\\
&$(\frac12,0)$&&&$-\frac83$&&$\frac{16}{3}$&&&$\frac{16}{3}$&$-\frac{8}{3}$&&5845(+)&2628(-)\\
\hline
\end{tabular}
\caption{The obtained CMI's and estimated masses for the case (3) in
units of MeV. The symbol ``+'' (``-'') after the tetraquark mass
means that $H_{CMI}>0$ ($H_{CMI}<0$).}\label{CMI3}
\end{table*}

\begin{table*}[h!]
\begin{tabular}{ccccccc|c||c}\hline
State&$(S_{qq\bar{q}},J)$&$\frac{C_{qq}}{m_q^2}$&$\frac{C_{qs}}{m_qm_s}$&$\frac{C_{ss}}{m_s^2}$&$\frac{C_{\bar{q}\bar{b}}}{m_qm_b}$&$\frac{C_{\bar{s}\bar{b}}}{m_sm_b}$ &Mass& Charm case\\
\hline $B_{\bar{s},15}$
&$(\frac12,1)$&$-4$&&&&$\frac43$&5770(+)&2567(+)\\
(5700)&$(\frac12,0)$&$-4$&&&&$-4$&5783(+)&2592(+)\\
\hline $B_{15}$, $B^{\prime\ell}$
&$(\frac12,1)$&$-4$&&&$\frac{4}{3}$&&5632(+)&2428(+)\\
(5560)&$(\frac12,0)$&$-4$&&&$-4$&&5639(+)&2450(+)\\
\hline $B^{\prime h}$
&$(\frac12,1)$&&$-4$&&&$\frac43$&5872(+)&2669(+)\\
(5840)&$(\frac12,0)$&&$-4$&&&$-4$&5885(+)&2694(+)\\
\hline $B_{s,15}$, $B^{\prime\ell}_s$
&$(\frac12,1)$&&$-4$&&$\frac43$&&5734(+)&2530(+)\\
(5700)&$(\frac12,0)$&&$-4$&&$-4$&&5741(+)&2551(+)\\
\hline $B^{\prime h}_s$
&$(\frac12,1)$&&&$-4$&&$\frac{4}{3}$&6068(+)&2865(+)\\
(5980)&$(\frac12,0)$&&&$-4$&&$-4$&6081(+)&2889(+)\\
\hline $B_{ss,15}$
&$(\frac12,1)$&&&$-4$&$\frac{4}{3}$&&5929(+)&2725(+)\\
(5840)&$(\frac12,0)$&&&$-4$&$-4$&&5936(+)&2747(+)\\
\hline
\end{tabular}
\caption{The obtained CMI's and estimated masses for the case (4) in
units of MeV. The symbol ``+'' (``-'') after the tetraquark mass
means that $H_{CMI}>0$ ($H_{CMI}<0$).}\label{CMI4}
\end{table*}

\section{Numerical Results}\label{sec3}

We present the calculated CMI's in Tables \ref{CMI1}-\ref{CMI4},
where a multiplicative factor $-\frac{3}{32}$ is implicitly assumed.
The mass below the state symbol in the first column in Tables
\ref{CMI1}-\ref{CMI4} is simply the sum of the mass of the four
quark without the chromomagnetic interaction. In order to estimate
the rough masses of the tetraquark states, one has to determine the
ten parameters in $H_{CMI}$. To do that, we calculate CMI's for the
ground state baryons and mesons and extract the values of the ten
parameters from the mass splittings. We collect the results in Table
\ref{para}. In the extraction of the parameters of the light
quark-antiquark interaction, we don not use the masses of the
pseudoscalar mesons as input since they are influenced by the chiral
symmetry and its spontaneous breaking. Here we adopt the light
quark-quark CMI values: $\frac{C_{q\bar{q}}}{m_q^2}=196$ MeV,
$\frac{C_{q\bar{s}}}{m_qm_s}=94$ MeV, and
$\frac{C_{s\bar{s}}}{m_s^2}=242$ MeV. The quark masses are taken
from our previous work $m_q=310$ MeV, $m_s=450$ MeV, $m_c=1430$ MeV,
and $m_b=4630$ MeV \cite{Liu:2004kd}. We present the estimated
tetraquark masses for the both bottom and charm cases in Tables
\ref{CMI1}-\ref{CMI4}. We mainly discuss the bottom tetraquarks
in the following. The charmed tetraquarks have very similar
features.

\begin{table*}[h!]
\begin{tabular}{cc|cc||c}\hline
Hadron& CMI & Hadron & CMI & Parameter (MeV)\\\hline
$N$&$-\frac34\frac{C_{qq}}{m_q^2}$ & $\Delta$& $\frac34\frac{C_{qq}}{m_q^2}$& $\frac{C_{qq}}{m_q^2}=196$\\
$\Sigma$&$-\frac14\frac{C_{qq}}{m_q^2}-\frac12\frac{C_{qs}}{m_qm_s}$&$\Sigma^*$&$\frac14\frac{C_{qq}}{m_q^2}+\frac12\frac{C_{qs}}{m_qm_s}$ &$\frac{C_{qs}}{m_qm_s}=94$ \\
$\Xi$&$-\frac12\frac{C_{qs}}{m_qm_s}-\frac14\frac{C_{ss}}{m_s^2}$&$\Xi^*$&$\frac12\frac{C_{qs}}{m_qm_s}+\frac14\frac{C_{ss}}{m_s^2}$
&$\frac{C_{ss}}{m_s^2}=242$\\
$\bar{D}$&$-\frac32\frac{C_{q\bar{c}}}{m_cm_q}$&$\bar{D}^*$&$\frac12\frac{C_{q\bar{c}}}{m_cm_q}$&$\frac{C_{q\bar{c}}}{m_cm_q}=72$\\
$\bar{D}_s$&$-\frac32\frac{C_{s\bar{c}}}{m_cm_s}$&$\bar{D}^*_s$&$\frac12\frac{C_{s\bar{c}}}{m_cm_q}$&$\frac{C_{s\bar{c}}}{m_cm_s}=72$\\
$\bar{B}$&$-\frac32\frac{C_{q\bar{b}}}{m_bm_q}$&$\bar{B}^*$&$\frac12\frac{C_{q\bar{b}}}{m_bm_q}$&$\frac{C_{q\bar{b}}}{m_bm_q}=23$\\
$\bar{B}_s$&$-\frac32\frac{C_{s\bar{b}}}{m_bm_s}$&$\bar{B}^*_s$&$\frac12\frac{C_{s\bar{b}}}{m_bm_q}$&$\frac{C_{s\bar{b}}}{m_bm_s}=25$\\
$\Sigma_c$&$\frac14\frac{C_{qq}}{m_q^2}-\frac{C_{qc}}{m_cm_q}$&$\Sigma_c^*$&$\frac14\frac{C_{qq}}{m_q^2}+\frac12\frac{C_{qc}}{m_cm_q}$ &$\frac{C_{\bar{q}\bar{c}}}{m_cm_q}=43$\\
$\Xi_c^\prime$&$\frac14\frac{C_{qs}}{m_qm_s}-\frac12\frac{C_{qc}}{m_cm_q}-\frac12\frac{C_{sc}}{m_cm_s}$ &$\Xi_c^*$&$\frac14\frac{C_{qs}}{m_qm_s}+\frac14\frac{C_{qc}}{m_cm_q}+\frac14\frac{C_{sc}}{m_cm_s}$&$\frac{C_{\bar{s}\bar{c}}}{m_cm_s}=49$\\
$\Sigma_b$&$\frac14\frac{C_{qq}}{m_q^2}-\frac{C_{qb}}{m_bm_q}$&$\Sigma_b^*$&$\frac14\frac{C_{qq}}{m_q^2}+\frac12\frac{C_{qb}}{m_bm_q}$ &$\frac{C_{\bar{q}\bar{b}}}{m_bm_q}=14$\\
$\Xi_b^\prime$&$\frac14\frac{C_{qs}}{m_qm_s}-\frac12\frac{C_{qb}}{m_bm_q}-\frac12\frac{C_{sb}}{m_bm_s}$ &$\Xi_b^*$&$\frac14\frac{C_{qs}}{m_qm_s}+\frac14\frac{C_{qb}}{m_bm_q}+\frac14\frac{C_{sb}}{m_bm_s}$&$\frac{C_{\bar{s}\bar{b}}}{m_bm_s}=26$\\
\hline
\end{tabular}
\caption{The CMI's for baryons and mesons to determine the
parameters in the model and obtained parameters. The parameter
$\frac{C_{\bar{s}\bar{b}}}{m_bm_s}=26$ MeV is estimated with the
mass difference $\Xi_b^*-\Xi_b^\prime\approx 30$ MeV taken from Ref.
\cite{bottombaryons}.}\label{para}
\end{table*}

In the case (1), one always gets $\langle H_{CMI}\rangle<0$. From
Table \ref{CMI1} and the value of the parameters, we notice that the
attraction mainly arises from the color-magnetic interaction between
the light quarks. The CMI's between quark and antiquark vanish in
the present case. Although the interaction between the antiquarks is
repulsive (negative values in Table \ref{CMI1}) for the vector case,
the resulting total interaction is still attractive.

The state $B_{s,\bar{6}}$ with $J=0$ corresponds to the $X(5568)$
although the mass is 51 MeV higher. Considering the model errors,
the tetraquark interpretation for the $X(5568)$ is favored. If this
state really exists, its tetraquark partners in Table \ref{CMI1}
should all exist. For example, the CMI for the $B^\ell$ state with
$J=0$ is about -160 MeV while the CMI is -80 MeV for $B_{s,\bar{6}}$
with $J=0$. The most attractive interaction occurs for
$B_{\bar{s},\bar{6}}$ (J=0) with $CMI=-166$ MeV. The larger CMI's
for $B^\ell$ and $B_{\bar{s},\bar{6}}$ lies in the fact that the
$\ell\ell$ ($\ell=u$ or $d$) interaction is stronger than the $\ell
s$ interaction while the interaction for the antiquarks does not
matter. We perform the calculation with flavor wave functions in the
SU(3) symmetry limit. Since there does not exist a symmetry
violating operator in the Hamiltonian, we get a degenerate result
for $B_{s,\bar{6}}$ and $B_{s,3}$. One expects a lower $B_{s,3}$
tetraquark state once a more realistic model is adopted.

In the case (2), the color-magnetic interaction for the $(0^+,1^+)$
doublet is always attractive. Although the quark-quark interaction
is not so attractive, the nonvanishing $q\bar{q}$ interaction
provides much stronger attraction. As a result, the obtained masses
for the tetraquark states with $S_{qq\bar{q}}=\frac12$ are lower
than those in the case (1). Now the $B_{s,\bar{6}}$ state with $J=1$
is close to the observed $X(5568)$, although the mass is 49 MeV
lower. In this case, the largest CMI (-300 MeV) occurs for the
$B^\ell$ state with $J=0$, again not for $B_{s,\bar{6}}$, which
indicates the existence of more stable non-strange tetraquark
states.

From Table \ref{CMI2}, our results indicate that although a ``good''
diquark is always used to discuss hadron spectrum, there may exist a
more attractive configuration for the multiquark states. More
tightly bound tetraquark states are possible with a not-so-good
diquark because of the existence of a light antiquark, which means
that a triquark structure seems more appropriate here.

In the case (3), the attractive color-magnetic interactions are
possible only for states with $S_{qq\bar{q}}=\frac12$. The
interaction for the light quarks is weakly repulsive but the
attractive interaction for the light quark and light antiquark may
result in $\langle H_{CMI}\rangle<0$. The mass of the state
corresponding to the $X(5568)$ is about 80 MeV higher than the
observation.

In the case (4), one always gets a positive $\langle
H_{CMI}\rangle$. The color-spin interaction for the light quarks is
always repulsive. The weaker attraction between the antiquarks does
not affect the final interaction significantly. There does not exist
a good candidate within the multiplet for the $X(5568)$ in this
case.

From the above analysis, we notice that the interaction among the
three light constituents dominantly affects the chromomagnetic
splitting and the stability of the tetraquark states. Moreover, the
state corresponding to $X(5568)$ is not the most tightly bound
tetraquark. A better triquark assumption may be more helpful for the
stability of a multiquark system than the ``good'' diquark
assumption, where the triquark satisfies the condition: (a) $qq$ in
$3_f$, $6_c$, and thus $S_{qq}=1$ and (b) $S_{qq\bar{q}}=\frac12$.
The triquark cluster has been proposed by Karliner and Lipkin in
discussion of the pentaquark $\Theta^+$ \cite{triquark}. Here, the
flavor representation of the triquark is not constrained to be
$\bar{6}_f$ only.

\section{Decay patterns}\label{sec4}

If the state $X(5568)$ does exist, the tetraquark interpretation is
favored and its partners should also exist. It is worthwhile to
search for such states in various channels. Now we turn to the decay
properties of these tetraquarks. We mainly discuss those states with
$\langle H_{CMI}\rangle<0$.

(a) States with quark content $\ell\ell\bar{s}\bar{b}$ ($I=1,0$):
$B_{\bar{s},\bar{6}}$, $B_{\bar{s},15}$. The possible strong decay
channel is $BK$. The $BK$ threshold is around 5774 MeV ($>H_0=5700$
MeV), so the temporary conclusion is that no strong decay channel is
allowed. In the charm case, although the $D_{\bar{s},\bar{6}}$ with
$J=1$ in the case (2) is above the $\bar{D}K$ threshold (2365 MeV),
the decay is forbidden by angular momentum conservation. In other
words, these tetraquark states are stable once produced.

(b) States with quark content $\ell\ell\bar{\ell}\bar{b}$
($I=\frac32,\frac12$): $B^\ell$, $B_{15}$, $B^{\prime\ell}$. Since
the $B\pi$ threshold (5420 MeV) and the $B^*\pi$ threshold (5465
MeV) are both smaller than $H_0=5560$ MeV, it is possible to find
tetraquark states in these channels. In the charmed case, the decay
into $\bar{D}\pi$ (2010 MeV) or $\bar{D}^*\pi$ (2150 MeV) is also
allowed.

(c) States with quark content $\ell s\bar{s}\bar{b}$ ($I=\frac12$):
$B^h$, $B^{\prime h}$. There are two types of strong decays: $B\eta$
and $B_sK$. The thresholds are 5827 MeV and 5861 MeV, respectively.
Only tetraquarks with repulsive color-magnetic interaction can
possibly decay. However, in the charmed case, the tetraquark states
decay into $\bar{D}\eta$ (2418 MeV), $\bar{D}^*\eta$ (2558 MeV),
$D_sK$ (2465 MeV), or even $D_s^*K$ (2607 MeV). It is interesting to
search for the charmed tetraquarks in these channels.

(d) States with quark content $s\ell\bar{\ell}\bar{b}$ ($I=1,0$):
$B_{s,6}$, $B_{s,3}$, $B_{s,15}$, $B_s^{\prime\ell}$. The $B\bar{K}$
threshold (5774 MeV) is larger than $H_0=5700$ MeV and the lower
tetraquark states can only decay into $B_s\pi$ (5506 MeV) and
$B_s^*\pi$ (5555 MeV). For the states below 5500 MeV, even the
former channel is closed. Although the $B_{s,3}$ tetraquark state
may be above the $B_s\pi$ threshold, its $B_s\pi$ decay mode is
forbidden by isospin conservation. In the charmed case, the obtained
scalar $D_{s,\bar{6}}(2397)$, $D_{s,3}(2397)$, $D_{s,15}(2426)$, and
$D_{s}^{\prime\ell}(2426)$ are all above the $\bar{D}\bar{K}$
threshold (2365 MeV) and the $D_s\pi$ threshold (2109 MeV). The
decay into these channels is allowed. The decay into $D_s^*\pi$ for
the spin-1 partners of these $D_{s}$ tetraquarks is also allowed. It
is very interesting to search for the possible charmed tetraquarks
in the isovector channels. That is, the charmed partners of
$X(5568)$ have more decay modes. Their ratios might be used to
identify the tetraquark nature, like the unconfirmed $D_{sJ}(2632)$
\cite{Liu:2004kd}.

(e) States with quark content $ss\bar{s}\bar{b}$ ($I=0$):
$B_s^{\prime h}$. The most possible channel is $B_s\eta$ whose
threshold is 5914 MeV ($<H_0=5980$ MeV). But, the strong decay for
the obtained tetraquarks with the attractive interaction is
kinematically forbidden. In the charmed case, however, both
$D_s\eta$ (2517 MeV) and $D_s^*\eta$ channels are kinematically
allowed.

(f) States with quark content $ss\bar{\ell}\bar{b}$ ($I=\frac12$):
$B_{ss,15}$. If such tetraquarks with repulsive interaction exist,
their decay into $B_s\bar{K}$ (5861 MeV) is probably kinematically
allowed. Similar observation holds for the charmed case.

Therefore, our numerical results indicate that the bottom tetraquark
states with the attractive interaction are stable. If one wants to
search for them through strong decay, the only possible channels are
$B\pi$, $B^*\pi$, or $B_s\pi$. On the other hand, the charmed
tetraquarks may be searched for in the channels $D\pi$, $D^*\pi$,
$D_s\pi$, $D_s^*\pi$, $D\eta$, $D^*\eta$, $D_s\eta$, $D_s^*\eta$,
$DK$, $D_sK$, or even $D_s^*K$.

\section{Conclusions}\label{sec5}

In this work, we have discussed the possible bottom tetraquark
states with the chromomagnetic interaction model. We have considered
four kinds of tetraquark structures according to the symmetry of the
two light quarks. We find that the observed $X(5568)$ can be
accommodated as a tetraquark candidate easily. Our analysis
indicates that the other possible bottom tetraquarks should also
exist and should be stable.

From the numerical results, we find that the stability of a
$qq\bar{q}\bar{b}$ tetraquark state is dominantly determined by the
interaction among the three light constituents. A triquark structure
may result in lower tetraquark masses than a diquark assumption.
Such a triquark structure is similar to that proposed in Ref.
\cite{triquark} but the flavor representation may also be $3_f$.

In the charmed case, there also exist many $qq\bar{q}\bar{c}$
tetraquark states. They have more strong decay channels than the
bottom partners. In particular, the experimental search for the
charmed partners of the $X(5568)$ are strongly called for in the
$D_s\pi$, $D_s^*\pi$, and isovector $\bar{D}\bar{K}$ channels.

\section*{Acknowledgments}
This project is supported by National Natural Science Foundation of
China under Grants No. 11275115, No. 11222547, No. 11175073, No.
11261130311 and 973 program. XL is also supported by the National
Youth Top-notch Talent Support Program ("Thousands-of-Talents
Scheme").

\end{document}